# Thermal and Transport Properties of U$_3$Si$_2$

Daniel J. Antonio[a*], Keshav Shrestha[a], Jason M. Harp[a], Cynthia A. Adkins[a], Yongfeng Zhang[a], Jon Carmack[a], and Krzysztof Gofryk[a*]

[a]Idaho National Laboratory, 2525 Fremont Ave, Idaho Falls, ID 83415, USA
*Corresponding author.
 E-mail addresses: daniel.antonio@inl.gov; krzysztof.gofryk@inl.gov

**Abstract**

We have studied U$_3$Si$_2$ by means of the heat capacity, electrical resistivity, Seebeck and Hall effects, and thermal conductivity in the temperature range 2-300 K and in magnetic fields up to 9 T. All the results obtained point to delocalized nature of 5$f$-electrons in this material. The low temperature heat capacity is enhanced ($\gamma_{el}$ ~ 150 mJ/mol-K$^2$) and shows an upturn in $C_p/T$ (T), characteristic of spin fluctuations. The thermal conductivity of U$_3$Si$_2$ is ~8.5 W/m-K at room temperature and we show that the electronic part dominates heat transport above 300 K as expected for a metallic system, although the lattice contribution cannot be completely neglected.

## 1. Introduction:

Binary uranium-silicon compounds have been extensively investigated for use as nuclear fuels in new generation reactors [1]. In particular, U$_3$Si$_2$ has become a material of interest for its potential as an accident tolerant fuel used in commercial light water reactors (LWR) [2]. The high-power densities in LWR cores make a pressing need for more accident tolerant fuel materials to replace existing Zr alloy clad uranium and plutonium oxide fuels. Extensive research has been performed on U$_3$Si$_2$ for use in dispersion plate type fuel for research reactors. This fuel performed well under these conditions, but in the anticipated LWR application, U$_3$Si$_2$ will be irradiated above the temperatures that were studied previously. Amorphization and swelling is expected to behave differently above 250°C, as discussed by Birtcher 1996 [3]. Several factors make U$_3$Si$_2$ appealing, including a higher uranium density than UO$_2$ (currently the most commonly used commercial fuel). An improved thermal conductivity over the comparatively poor UO$_2$ is also of particular note, as better thermal properties can contribute to power uprate potential from higher linear heat generation rates in the fuel and smaller temperature gradients during reactor start up could reduce cracking of fuel pellets. A higher thermal conductivity is also beneficial during some accident scenarios.

Although much effort has been dedicated to studying the thermo-physical properties of U$_3$Si$_2$ at high temperatures, there is only limited data available at lower temperatures. The operating temperatures in nuclear reactors are high (~1000 K), however, many important physical characteristics such as the effect of electronic correlations and/or impact of defects and other degrees of freedom on the electrical and heat transport in nuclear materials are all emphasized at moderate or low temperatures. Therefore, in order to better understand the nature of the 5f-electrons and mechanisms that govern electrical and heat transport in this important technological material, and to accurately model this compound at all relevant temperatures, these effects must be quantified.

The minimum U-U distance in $U_3Si_2$ is 3.32 Å [4], which is below the Hill limit [5], leading to the expectation of magnetic moments being suppressed. In fact, previous magnetic and electrical studies suggest a large degree of 5*f*-electron delocalization in $U_3Si_2$. The low temperature magnetic susceptibility shows little temperature dependence and no transition to a magnetically ordered state, characteristic of Pauli paramagnetism [4] [6] [7]. In addition, a broad maximum in magnetic susceptibility around 300 K and a characteristic *s*-shaped $\rho$(T) curve [8] suggest the presence of spin fluctuations in the uranium 5*f* moments in $U_3Si_2$ [9]. The dependence between Knight Shift and spin relaxation rate values obtained from $Si^{29}$ NMR compared to the susceptibility are also similar to prototypical spin-fluctuator $UAl_2$ [9]. A recent study on hydrogenated $U_3Si_2H_{1.8}$, which has the same crystal structure as $U_3Si_2$ with expanded lattice parameters, also suggested the presence of spin fluctuations [7].

In this work, we have investigated the low temperature thermal properties of $U_3Si_2$, as manufactured using industry scalable techniques for testing in future nuclear applications under LWR conditions. This gives the opportunity to characterize the material in the form it might be used in real applications. All results, especially enhanced low temperature heat capacity, characteristic behavior of magnetoresistivity, and thermoelectric power, are consistent with delocalized nature of 5*f*-electrons in this material. The low temperature heat capacity data obtained for $U_3Si_2$ is consistent with a spin fluctuation model. We show that at room temperature the thermal conductivity of $U_3Si_2$ is dominated by the electronic part, although a smaller lattice contribution still remains.

**2. Experimental Details:**
Polycrystalline samples of $U_3Si_2$ were prepared by arc-melting stoichiometric amounts of elemental U and Si. The arc melted ingots were then comminuted into powder, pressed, and sintered into pellets, as detailed in ref. [10]. Powder diffraction from that work confirmed that the structure was tetragonal (space group P4/mbm) with lattice parameters similar to those previously reported in literature [4], though minor phases of USi and $UO_2$ were seen in the pattern. Though X-ray diffraction of the sample showed a presence of a small amount of $UO_2$ phase, the strong magnetic feature near 30 K in $UO_2$ $C_p$ was not observed [11] consistent with the very small amount of this material present in the samples studied. Helium pycnometry showed pellet density of 11.54±0.06 g/cm³ (~95% theoretical maximum). Samples for this work were sectioned from these sintered pellets. A sketch of the crystal structure of $U_3Si_2$ is shown in the inset of Figure 1. The thermal conductivity, resistivity, Hall effect, and heat capacity measurements were done in a DynaCool Quantum Design Physical

Property Measurement System (PPMS) system equipped with a 9 T superconducting magnet. The sample configuration for resistivity and Hall effect measurements was a typical 4-lead *ac* method using platinum wire leads connected with Epo-Tek silver-filled H20E epoxy. Typical errors in the resistivity and Hall effect measurements are ~2%, mainly due to uncertainty in measurement of sample form factor (approximately 2 x 1 x 1 mm bars) and in case of heat capacity it is less than 3%. The thermal conductivity and thermopower measurements were done in continuous heating mode of the TTO option using a pulse-power steady state method in the PPMS as the sample was ramped from 2 Kelvin to 300 Kelvin at 0.25 K/min. Errors using the TTO option are largest at higher temperatures due to corrections for heat lost from thermal radiation, reaching ~5% near room temperature. In order to obtain the high temperature thermal conductivity, a combination of laser flash analysis (LFA) and differential scanning calorimetry (DSC) was used to measure the thermal diffusivity ($\alpha$) and specific heat capacity ($c_p$), respectively, of samples from this same batch from room temperature to 1000 K under gettered high purity (UHP) argon. The LFA measurements were performed on a Netzsch LFA 457 instrument following ASTM E 1461-13 [12]. The DSC measurements were performed on a Netzsch DSC 404 F1 instrument according to ASTM E1269-11 [13]. Values for thermal conductivity were calculated according to $\kappa(T) = \alpha_T c_p \rho_D$, where $\kappa$ is the thermal conductivity, $\alpha_T$ is thermal diffusivity, $c_p$ is the specific heat capacity and $\rho_D$ is the mass density as a function of temperature, resulting in maximum errors of ~5%.

## 3. Results and Discussion:

*3.1 Electrical Resistivity and Magnetoresistivity*

The temperature dependence of the electrical resistivity, $\rho$, of $U_3Si_2$ is shown in Figure 1. The resistivity decreases with decreasing temperatures typical for metallic systems, exhibiting an inflection point at approximately 50 *K*. This "*s*-shaped" behavior is characteristic of strongly correlated 4 and 5*f*-electron systems, especially spin fluctuators [14] [15]. For comparison, in the inset, we have also included the temperature dependencies of the electrical resistivity of $U_3Si_2$ from previous studies [6] [8] [16]. As seen from the figure, our room temperature resistivity agrees with some previous reports [16], but is lower than others [6][8]. The difference in absolute magnitudes of the resistivity found in $U_3Si_2$ might suggest that the electrical transport is sensitive to details of the atomic (for instance, off-stoichiometry) or electronic disorder in this material. In order to investigate this behavior in more detail electrical measurements on single crystalline materials are required. To study the influence of magnetic field on transport properties in $U_3Si_2$, we measured transverse magnetoresistivity in magnetic fields as strong as 9 T. In Figure 2, we plot the temperature dependence of magnetoresistivity, defined as *MR = [ρ(H)-ρ(0)]/ρ(0)*. As seen from the figure, the magnetoresistance is typical of non-magnetic metals, showing a small positive value that increases with increasing applied field [17] [6]. *MR* measured in *H* = 9 T is very small, of the order of 0.45 % at 2 K. With increasing temperature *MR* decreases and saturates at ~0.05 % above 30 K. An anomaly can be seen at approximately 20 K, the origin of which is not clear at the moment and is a subject of ongoing investigations. The inset in Figure 2 shows several isotherms of the magnetoresistivity (MR) of $U_3Si_2$ as a function of applied magnetic field. Interestingly, both *MR(T)* and *MR(H)* display similar behavior to $TiBe_2$ and $UAl_2$, where it was shown that it is related to ordinary magnetoresistivity due to the orbital motion of the electrons in magnetic field and due to spin fluctuations [18].

*3.2 Seebeck and Hall effect*

The Seebeck coefficient, *S*, as a function of temperature is shown in Figure 3a. The value at room temperature of about 14.3 µV/K is larger than observed in simple metals such as Cu, Au, or Ag and its absolute magnitude is much closer to that of Pd [19]. With lowering temperature, the thermopower smoothly decreases with decreasing temperature down to 2 *K*, our lowest temperature measured. The overall temperature dependence of the Seebeck coefficient of $U_3Si_2$ is similar to some other U-based correlated materials [20] [21] [22]. Accordingly, *S(T)* for $U_3Si_2$ was analyzed in the framework of a model that takes into account scattering conduction electrons by a 5*f* quasiparticle band of a Lorentzian form [20]. In this so-called two-band model the thermoelectric power is given by:

$$S(T) = \frac{AT}{(B^2 + T^2)}, \text{ where } A = \frac{2\varepsilon}{|e|} \text{ and } B^2 = \frac{3(\varepsilon^2 + \Gamma^2)}{\pi^2 k_B^2}. \quad (1)$$

The symbol *ε* denotes the energy position of the 5*f* band with respect to the Fermi level and *Γ* stands for its bandwidth. As shown in Fig. 3a, above 170 K the model provides a good approximation of the experimental results of $U_3Si_2$ ($r^2$ = *0.997)* with parameters *ε* = 7.8 meV, *Γ* = 76 meV. These values are similar to those derived for several U, Np, or Pu-based intermetallics with strong electronic correlations [20] [21] [22] [23] [24]. In addition, assuming a single-band model and scattering from atomic disorder being dominant at high temperature, the Fermi energy $\varepsilon_f$ can be approximated by $\varepsilon_f$ = $k_B^2$ $\pi^2$ *T*/3|*e*|*S* [25]. This gives a value of $\varepsilon_f$ = 0.513 eV, and an estimate for the effective carrier concentration $n_s$ ~$10^{21}$ $cm^{-3}$. As seen in Figure 3a, below 170 K the thermopower of $U_3Si_2$ shows some deviation from the model used. This might be indicative of the presence of more complex electronic structure in this material. Also, other contributions to the thermoelectric power such as phonon drag effect [26] might contribute to the total S(T) in $U_3Si_2$.

The temperature dependence of the Hall coefficient, $R_H$, is also shown in Figure 3a. The value of the coefficient is negative over the entire temperature range, suggesting that electrons with high mobility are dominant charge carriers. The *$R_H(T)$* shows little temperature dependence and its value is of the order of -5 ×$10^{-4}$ $cm^3$/C, one order of magnitude larger than $R_H$ of Copper (~-5 ×$10^{-5}$ $cm^3$/C [27]). Once $R_H$ is determined, the charge carrier density (*n*) can be estimated as *n = 1/($R_H$e)*, as seen in

Figure 3b. At 250 K, this corresponds to 1.4 mobile charge carriers per formula unit. Similarly, using the measured electrical resistivity ($\rho$), the charge carrier mobility ($\mu$) can be determined using the formula $\mu = R_H/\rho$, as shown in Figure 3b, giving ~15 cm$^2$V$^{-1}$s$^{-1}$ at 250 K. The so-obtained carrier concentrations and motilities can be compared to other strongly correlated metals using their conventional temperature independent R$_H$ in the limit much larger than the Kondo temperature, such as CeNiGe$_3$ with $n$ = 6.7 ×10$^{-21}$ carriers/cm$^3$ (0.65 carriers/f.u.) and $\mu$ ~ 9 cm$^2$V$^{-1}$s$^{-1}$ [28], UGe$_2$ with $n$ = 6.6 ×10$^{-21}$ carriers/cm$^3$ (0.4 carriers/f.u.) and $\mu$ ~ 2.5 cm$^2$V$^{-1}$s$^{-1}$ [29], and U$_2$Zn$_{17}$ with $n$ = 14 ×10$^{-21}$ carriers/cm$^3$ (2.9 carriers/f.u.) and $\mu$ ~ 3.4 cm$^2$V$^{-1}$s$^{-1}$ [30].

The negative Hall coefficient together with positive Seebeck coefficient, indicate that a simple free electron model cannot be used to describe the electronic properties of this system. DFT calculations using spin-orbit coupling and an on-site coulomb correction [31] [32] indicate that besides uranium 5$f$-electrons there are also several overlapping states, in particular 6$d$ and 3$p$ at the Fermi level. Interplay between conventional and extraordinary components with differing electron-like or hole-like character could lead to anomalous values for $R_H$ [33].

*3.4 Heat Capacity*

The temperature dependence of the heat capacity ($C_p$) of U$_3$Si$_2$ is shown in Figure 4. The value of $C_p$ at 300 K is 149 J/mol-K, which is consistent with previous high temperature studies [34]. This is, however, higher than the theoretical Dulong-Petit limit for the phonon specific heat contribution, $3Rn$ = 124.71 J/mol-K, where $R$ is the Gas constant and $n$ stands for number of atoms in the formula unit. This suggests that there may be a large electronic component to the specific heat, even at room temperature. The inset of Figure 4 shows the low temperature heat capacity depicted in a $C_p/T$ vs. $T^2$ plot. The upturn at low temperature is a feature common to strongly correlated compounds that exhibit spin fluctuations [35] [36] [37] [38]. Recent heat capacity measurements of U$_3$Si$_2$H$_{1.8}$ also show an upturn in the low temperature heat capacity [10]. In this class of materials, the low temperature heat capacity can be expressed as [39]:

$$\frac{C_p}{T}(T) = \gamma + \beta T^2 + \frac{\alpha \gamma}{T_{sf}^2} T^2 \ln\left(\frac{T}{T_{sf}}\right), \quad (2)$$

where $\gamma$ is the renormalized value of the Sommerfeld coefficient that is proportional to electronic density of states at the Fermi level and consequently the effective mass of carriers, the $\beta$ term is the lattice contribution related to Debye temperature by:

$$\theta_D = \left(\frac{12R\pi^4 n}{5\beta}\right)^{\frac{1}{3}}, \quad (3)$$

the $\alpha$ parameter is proportional to inter-spin Coulomb repulsion, density of states, and Stoner exchange enhancement [39], and $T_{sf}$ stands for the temperature of spin fluctuations. A least-square fit to the data below ~10 K gives; $\gamma$ = 149.7 mJ/mol-K$^2$, $\beta$ = 1.27 x 10$^{-3}$ J/mol-K$^4$, $\alpha$ = 170.7 J/mol-K$^6$, $T_{sf}$ = 207.8 K and is shown by the solid line in the inset of figure 5. An estimation of the Debye temperature can be obtained from $\beta$ using equation 3 to be $\Theta_D$ = 197 K, which agrees with previous estimates [34] [40] [41].

*3.5 Thermal Conductivity*

Figure 5 shows the temperature dependence of the thermal conductivity ($\kappa$) of U$_3$Si$_2$. As seen from the figure a monotonic increase over the temperature range measured, rising to a value of 8.4 W/m-K at 300 *K*. This is in agreement with previously published high temperature studies [34] [42]. In metals and intermetallics the thermal conductivity is governed by contributions coming from electrical carriers and from lattice vibrations (phonons) and may be approximated by $\kappa(T) = \kappa_{el}(T) + \kappa_{ph}(T)$. The contribution of the electronic thermal conductivity ($\kappa_{el}$) can be estimated from the Wiedemann-Franz law:

$$\kappa_{el}(T) = \frac{LT}{\rho(T)}, \qquad (4)$$

which relates $\kappa_{el}$ to the electrical resistivity $\rho(T)$ through the Lorenz number, *L* = 2.44 ×10$^{-8}$ WΩ/K. In many cases the Lorenz number can be considered as temperature independent over a wide range of temperatures and materials [43]. In strongly correlated materials, though, some deviations from the ideal value of *L* have been observed at low temperatures (below 100 K), especially in systems with non-Fermi liquid behavior or near a quantum critical point [46]. By subtracting $\kappa_{el}$ from the total, it can be seen from Fig. 5 that the so-obtained lattice thermal conductivity $\kappa_{ph}$ is relatively small reaching ~2 W/mK at room temperature. In order to extend the thermal conductivity determination to higher temperature range we have performed high temperature laser-flash measurements performed on the same samples as used in the low temperature studies (see the inset in Fig.5). It is worth noting that the measured conductivity is consistent with previous experimental results for U$_3$Si$_2$ in the high temperature regime [34] [44]. As shown in the inset of Fig. 5, at about room temperature the results are very close, and the trends with respect to increasing temperature are nearly identical. Agreement between all these values indicates that presence of minor phases in U$_3$Si$_2$ do not significantly alter the total thermal conductivity of the samples. In the context of the heat transport performance of U$_3$Si$_2$ it is worthwhile to compare the measured lattice thermal conductivity to the theoretically achievable minimum of the phonon contribution. The latter may be derived from the expression [45]:

$$\kappa_{L_{min}} = \left(\frac{3n}{4\pi}\right)^{\frac{1}{3}} \frac{k_B^2 T^2}{\hbar \theta_D} \int_0^{\frac{\theta_D}{T}} \frac{x^3 e^x}{(e^2-1)^2} dx \qquad (5)$$

In this model no distinction is made between the transverse and longitudinal acoustic phonon modes. The result obtained for $U_3Si_2$ ($\theta_D$ = 197 K and n = 4.75$^{28}$ m$^{-3}$) is shown in Fig. 5 by the solid line.

## 4. Summary and Conclusions

To summarize, we present the heat capacity, electrical resistivity, Seebeck and Hall effects, and thermal conductivity of $U_3Si_2$. The measurements were performed in wide temperature and magnetic field ranges (2-300 K and in magnetic fields up to 9 T). No magnetic ordering has been found down to 2 K and all the results obtained, especially small magnetoresistivity, large low-temperature heat capacity, and characteristic dependence of the Seebeck coefficient, point to delocalized nature of 5*f*-electrons in this material. The low temperature heat capacity is enhanced ($\gamma_{el}$ ~ 150 mJ/molK$^2$) and shows an upturn in $C_p/T$(T), characteristic of systems with spin fluctuations. The thermal conductivity of $U_3Si_2$ is ~8.5 W/m-K at 300 K and is governed by electronic and lattice contributions. The lattice part of the total thermal conductivity is relatively small in $U_3Si_2$, with electrons dominating heat transport above 300 K. This knowledge of the details of the heat transport in $U_3Si_2$ will be useful for researchers working on modeling and simulations of this new advanced fuel. Future measurements on single crystal samples would be useful to further study these properties in a more idealized case with better structural and atomic order.

**Acknowledgments:**

This work was supported by Advanced Fuel Campaign and the DOE's Early Career Research Program. Y.Z. thanks the support of the Accident Tolerant Fuel high-impact-problem project under the DOE NEAMS program.

**References:**

[1] Y.S. Kim, "Uranium intermetallic fuels (U–Al, U–Si, U–Mo)" in *Comprehensive Nuclear Materials vol. 3*, R. Konings, Ed., Elsevier Ltd, 2012, pp. 391-420.

[2] S.J. Zinkle, K.A. Terrani, J.C. Gehin, L. J. Ott, L.L. Snead, "Accident tolerant fuels for LWRs: A perspective," J. Nucl. Mater. 448 (1–3) (2014) 374–379.

[3] R.C. Birtcher, J.W. Richardson, M.H. Mueller, J. Nucl. Mater. 230 (1–3) (1996) 158-163.

[4]  K. Remschnig, T. Le Bihan, H. Nobl, and P. Rogl, "Structural Chemistry and Magnetic Behavior of Binary Uranium Silicides," J. Solid State Chem. 13 (1992) 391-399.

[5]  H.H. Hill, "Plutonium and Other Actinides," in *Proceedings of 4th International Conference on Plutonium and other Actinides*, Santa Fe (1970).

[6]  T. Miyadai, H. Mori, T. Oguchi, Y. Tazuke, H. Amitsuka, T. Kuwai and Y. Miyak, "Magnetic and electrical properties of the U-Si system (part II)," J. Magn. Mag. Mater. 104-107 (1992) 47-48.

[7]  S. Mašková, K. Miliyanchuk, L. Havela, "Hydrogen absorption in $U_3Si_2$ and its impact on electronic properties," J. Nucl. Mater. 487 (2017) 418-423.

[8]  T. Miyadai, H. Mori, Y. Tazuke and T. Komatsubara, "Magnetic and electrical properties of the U-Si system," J. Magn. Mag. Mater. 90-91 (1990) 515-516.

[9]  B. Nowak, O.J. ZogaI, K. Niediwiedi, R. Troc, K. Wochowski and Z. Zolnierek, "29Si NMR and magnetic susceptibility of U3Si2," Physica B 192 (1993) 213-218 https://doi.org/10.1016/0921-4526(93)90022-X.

[10] Jason M. Harp, Paul A. Lessing, Rita E. Hoggan, "Uranium silicide pellet fabrication by powder metallurgy for accident tolerant fuel evaluation and irradiation," J. Nucl. Mater. 466 (2015) 728-738, https://doi.org/10.1016/j.jnucmat.2015.06.027

[11] G. D. Khattak, "Specific heat of uranium dioxide (UO2) between 0.3 and 50 K," Phys. Stat. Sol. 75,1 (1983) 317-321

[12] *ASTM E1461-13, Standard Test Method for Thermal Diffusivity by the Flash Method, ASTM International, West Conshohocken, PA, 2013, www.astm.org.*

[13] *ASTM E1269-11, Standard Test Method for Determining Specific Heat Capacity by Differential Scanning Calorimetry, ASTM International, West Conshohocken, PA, 2011, www.astm.org.*

[14] H. J. Van Ddaal and K. H. J. Buschow, "Kondo Effect in Some Intermetallic Compounds of Ce," Phys. Stat. Sol. (a) 3, 853 (1970).

[15] J.M. Lawrence, P.S. Riseborough, R.D. Parks, Rep. Prog. Phys 44, 1 (1981).

[16] W. Martienssen, H. Warlimont, *Springer Handbook of Condensed Matter and Materials Data*, Springer Science, 2006, p. 474.

[17] T. McGuire, R. Potter, "Anisotropic Magnetoresistance in Ferromagnetic 3d Alloys," IEEE Mag. Soc. 11,4 (1975).

[18] J. M. van Ruitenbeek, Phys. Rev. B 34 (1986) 8507.

[19] N. Cusack and P. Kendall, "The Absolute Scale of Thermoelectric Power at High Temperature," Proc. Phys. Soc. 72 (5) (1958) 898.

[20] U. Gottwick, K. Gloss, S. Horn, F. Steglich, N. Grewe,  J. Magn. Magn. Mater. 47-48 (1985) 536.

[21] Y. Bando, T. Suemitsu, K. Takagi, H. Tokushima, Y. Echizen, K. Katoh, K. Umeo, Y. Maeda, T. Takabatake, " J. Alloys Compd. 1 (2000) 313".

[22] K. Gofryk, D. Kaczorowski, A. Czopnik, Solid state communications 133 (2005) 625.

[23] M. Szlawska, K. Gofryk, J.C. Griveau, E. Colineau, P. Gaczynski, R. Jardin, R. Caciuo, D. Kaczorowski, Phys. Rev. B 85 (2012) 134443.

[24] K. Gofryk, J.C. Griveau, P. S. Riseborough, T. Durakiewicz, Phys. Rev. B 94 (2016) 195117.

[25] R.D. Bernard, *Thermoelectricity in Metals and Alloys*, Taylor and Francis, London, 1972.

[26] F.J. Blatt, P.A. Schroeder, C.L Foiles and D. Greig, *Thermoelectric power of metals*, Plenum Press, New York, 1976.

[27] N.J. Simon, E.S. Drexler, and R.P. Reed, "Properties of Copper and Copper Alloys at Cryogenic Temperatures" ( NIST MN 177).

[28] A. P. Pikul, D. Kaczorowski, T. Plackowski, A. Czopnik, H. Michor, E. Bauer, G. Hilscher, P. Rogl, Yu. Grin, "Kondo behavior in antiferromagnetic $CeNiGe_3$," Phys. Rev. B 67 (2003) 224417.

[29] V. H. Tran, S. Paschen, R. Troć, M. Baenitz, and F. Steglich, "Hall effect in the ferromagnet $UGe_2$," Phys. Rev. B 69 (2004) 195314 .

[30] T. Siegrist, M. Olivier, S.P. McAlister, R.W. Cochrane, "Magnetotransport in the heavy-fermion compound $U_2Zn_{17}$," Phys. Rev. B 33 (1986) 4370.

[31] T. Wang, N. Qiu, X. Wen, Y. Tian, J. He, K. Luo, X. Zha, Y. Zhou, Q. Huang, J. Lang, S. Du, "First-principles investigations on the electronic structures of $U_3Si_2$," Journal of Nuclear Materials 469 (2016) 194-199.

[32] M.J. Noordhoek, T.M. Besmann, D. Andersson, S.C. Middleburgh, A. Chernatynskiy, "Phase equilibria in the U-Si system from first-principles calculations," J. Nucl. Mater. 479 (2016) 216-223.

[33] V.H. Tran, "Hall effect in strongly correlated electron systems," Materials Science Poland 24 (2006) 699.

[34] J.T. White, A.T. Nelson, J.T. Dunwoody, D.D. Byler, D.J. Safarik, K.J. McClellan, J. Nucl. Mater. 464 (2015) 275–280.

[35] A. de Visser, J. J. M. Frame, A. Menovskyt, and T. T. M. Palstrag, "Spin fluctuations and superconductivity in UPt," J. Phys. F: Met. Phys. 14(1984) L191-LI96.

[36] G.R. Stewart, "Heavy-fermion systems," Rev. Mod. Phys. 56 (1984) 755 .

[37] R. J. Trainor, M. B. Brodsky, and H. V. Culbert, "Specific Heat of the Spin-Fluctuation System $UAl_2$," Phys. Rev. Lett. 34, 16 (1975) 1019.

[38] G. R. Stewart, A. L. Giorgi, B.L. Brandt, S.Foner, and A.J. Arko, "High-field specific heat of the spin-fluctuation system $UAl_2$," Phys Rev B 28, 3 (1983) 1524.

[39] R. J. Trainor, M. B. Brodsky, and L. L. Isaacs , "Calorimetric evidence for spin fluctuations in $UAl_2$," AIP Conference Proceedings 24 (1975) 220.

[40] J.T. White, A.T. Nelson, D.D. Byler, D.J. Safarik, J.T. Dunwoody, K.J. McClellan, J. Nucl. Mater 456 (2015) 442-448.

[41] M. Rosen, Y. Gefen, G. Kimmel, A. Halwany, Phil. Mag. 28 (5) (1973) 1007-1014.

[42] H. Shimizu, "The Properties and Irradiation Behavior of U3Si2," Tech. Rep. NAA- SR-10621, Atomics International (1965).

[43] C. Kittel, *Introduction to Solid State Physics*, 5th Ed., 1976, p. 178.

[44] J.T. White, A.T. Nelson, J.T. Dunwoody, D.J. Safarik, K.J. McClellan, Corrigendum to "Thermophysical properties of U3Si2 to 1773 K", J. Nucl. Mater. 484 (2017) 386–387.

[45] D. G. Cahill and R. O. Pohl, Solid State Commun. 70 (1989) 927.

[46] C. Bonnelle and N. Spector, *Rare-Earths and Actinides in High Energy Spectroscopy*, Springer, Dordrecht (2015) pp. 79-157.

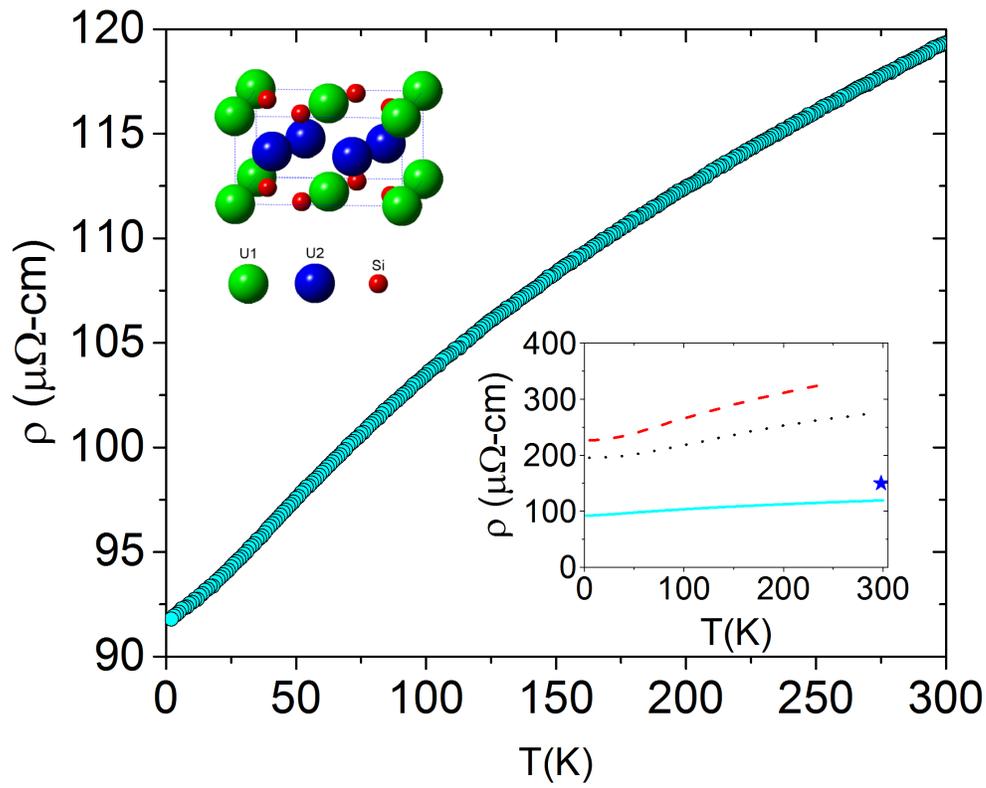

Figure 1. The temperature dependence of the electrical resistivity of $U_3Si_2$. Left inset: the tetragonal unit cell of $U_3Si_2$, showing the 2 unique uranium sites. Right inset: electrical resistivity of $U_3Si_2$ from this work (solid cyan line) compared to that from ref. [6] (dotted black), ref. [8] (dashed red) and the single room temperature value from ref. [14] (blue star).

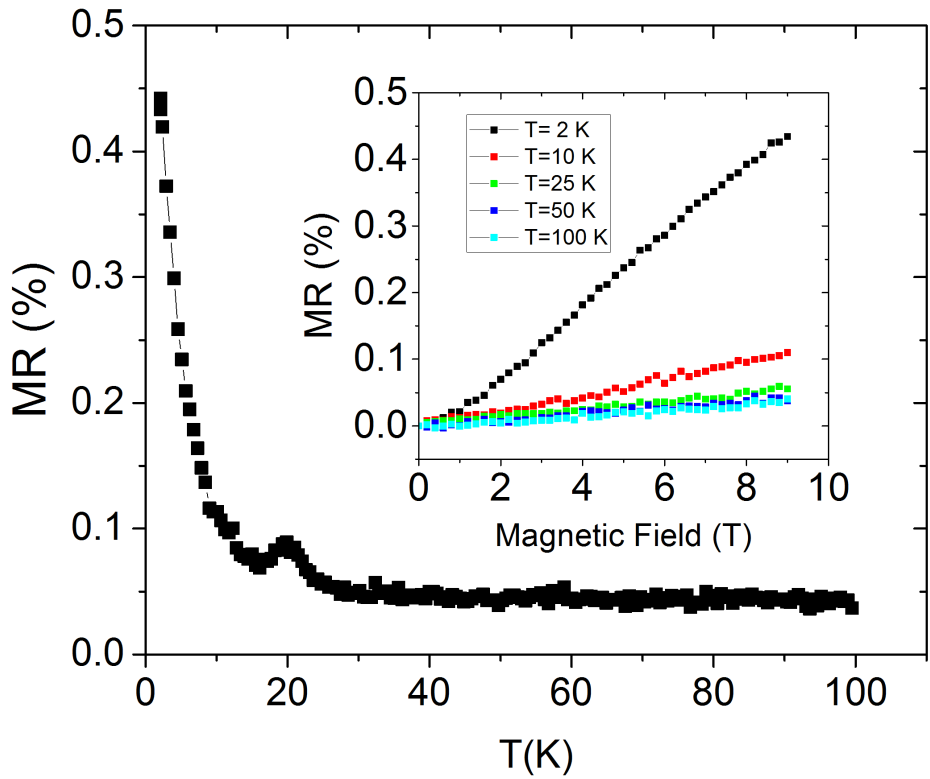

Figure 2. The temperature dependence of magnetoresistivity (MR) measured from 2 to 100 K under applied magnetic field of 9 T. Inset: The field dependence of the magnetoresistivity of $U_3Si_2$ taken at several temperatures.

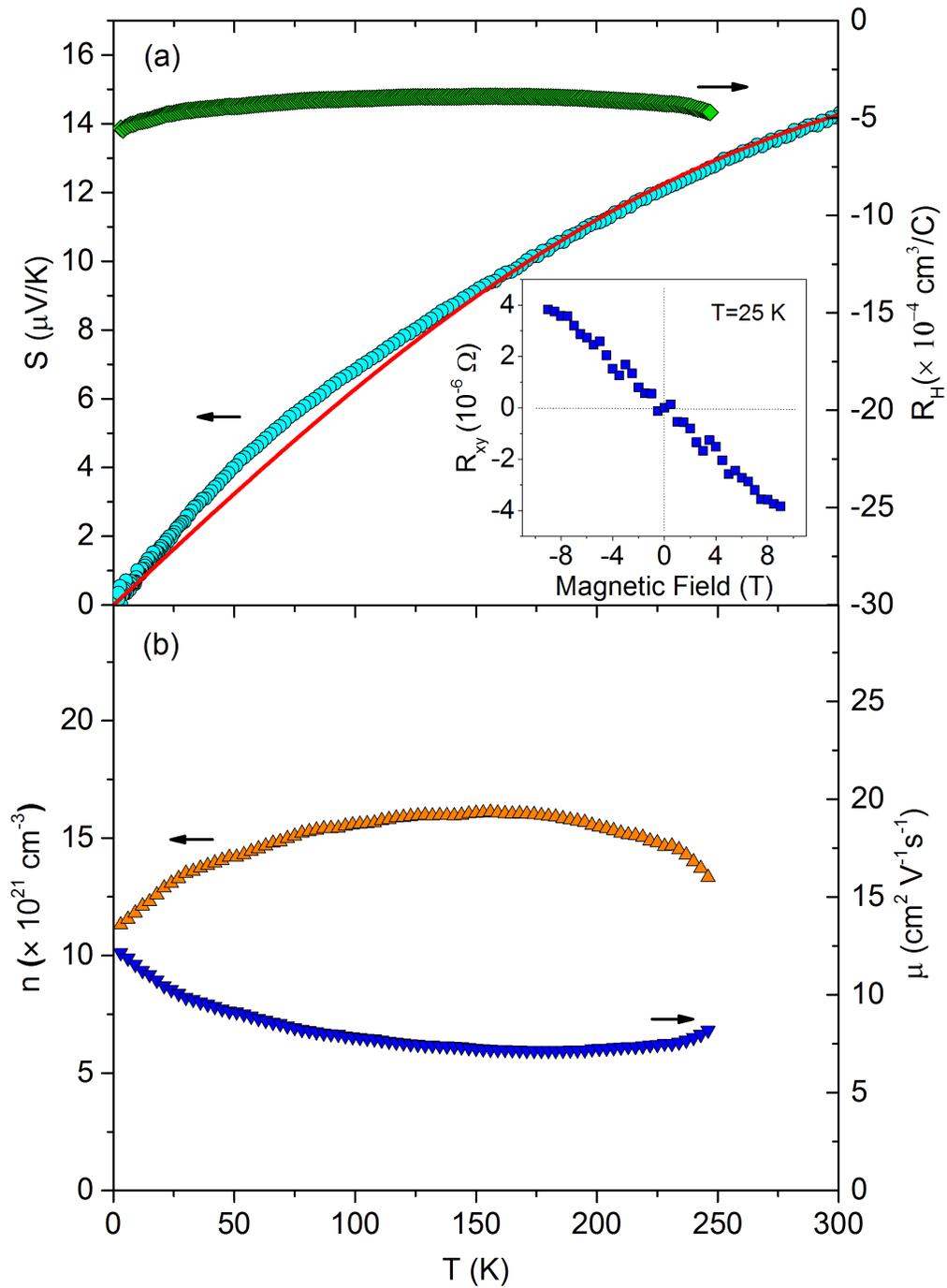

Figure 3. (a) The temperature dependence of the Seebeck coefficient (left scale) and the Hall coefficient (right scale) of $U_3Si_2$. Inset: The Hall resistance as a function of applied magnetic field measured at 25 K, showing linear dependence in this range. (b) The temperature dependence of carrier concentration (*n*) (left scale) and carrier mobility (*μ*) (right scale) of $U_3Si_2$.

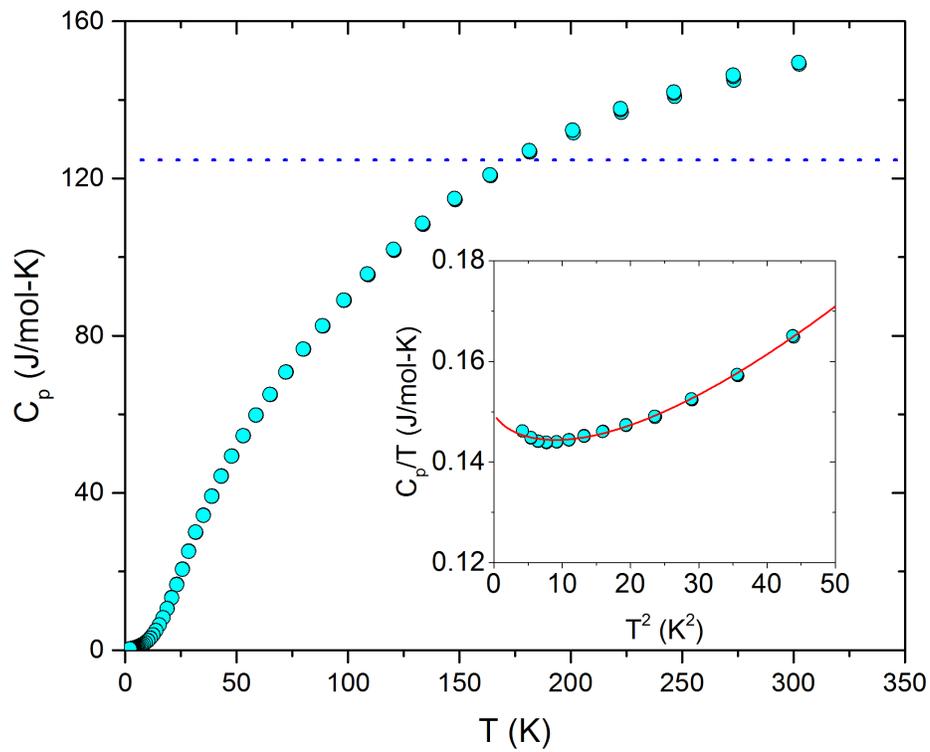

Figure 4. The temperature dependence of the heat capacity of $U_3Si_2$. The dotted line marks the theoretical Dulong-Petit value. Inset: the low temperature part of the heat capacity of $U_3Si_2$. The solid line is a fit of the equation 2 as described in Section 3.4.

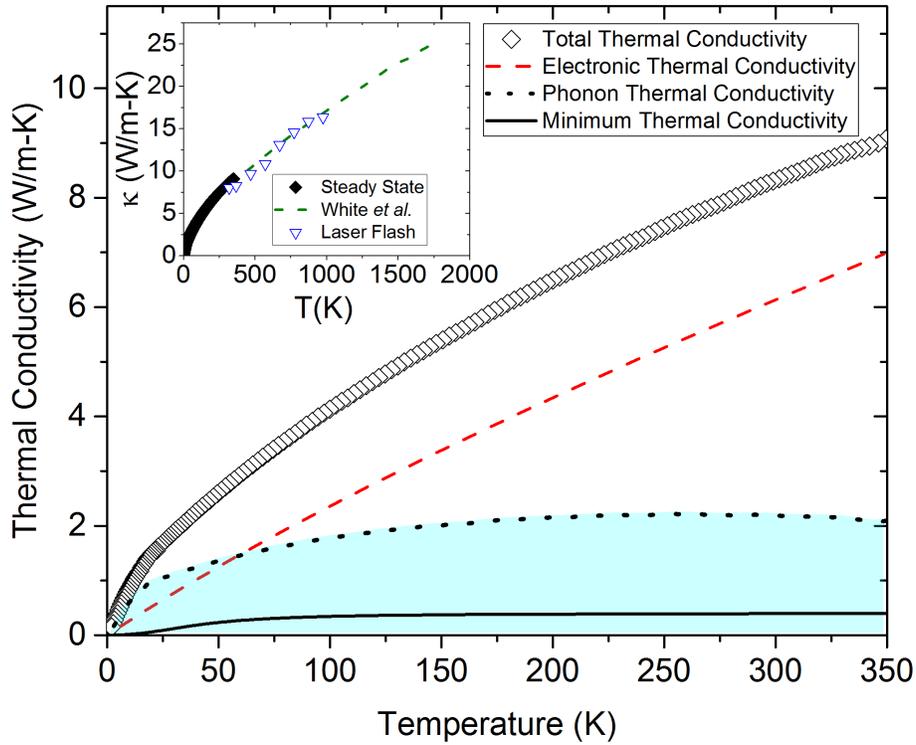

Figure 5. The temperature variation of the thermal conductivity of $U_3Si_2$. The dashed line is the electronic component calculated using equation 4. The dotted line represents the lattice part of the thermal conductivity of $U_3Si_2$, while the mimimum thermal conductivity from equation 5 is a solid line. Inset: the thermal conductivity of $U_3Si_2$ from 2 up to 1800 K. In this inset we include the low temperature measurements (steady-state), a laser flash study performed at INL on the same materials (laser flash) and previous laser flash measurements performed at LANL by White at al [44].